\documentclass[12pt]{article}
\usepackage{graphicx}
\usepackage{longtable}
\usepackage {amssymb}
\usepackage {amsmath}
\begin{document}
\title
{Elementary quantum cloning machines}
\author{V.N.Dumachev}
\maketitle

\begin{abstract}
The task of reception of a copy of an arbitrary quantum state with
use of a minimum quantity of quantum operations is considered.
\end{abstract}

\section{Introduction}

It is known, that an arbitrary quantum state cannot be copied
perfectly [1]. However V.Buzek and M.Hillery in work [2] offered a
universal quantum cloning machine (UQCM), allowing to create 2
identical qubits from 1 qubit. Two output qubits are a copy of
each other, but they are not a copy of an initial quantum state,
and are similar to it only in $5/6 \cong 0.83$. Universality of
the UQCM is that it clones any quantum state with identical
accuracy. In the present work some variants of not universal QCM
are considered. Their lack is that they clone qubits non-uniformly
depending on their state. In a final section  the calculation
scheme  of UQCM from  the general principles is submitted which
includes known results [3,4] as special cases. It is shown, how
one can  choose optimum by quantity of used quantum operations
UQCM from given ones.

\subsection{Brief theoretical data}

Any quantum state
\begin{equation}
\left| {\psi _{0}}\right\rangle =\alpha \left| {0}\right\rangle
+\beta \left| {1}\right\rangle   \label{1.1}
\end{equation}
\noindent is represented by a point on Bloch sphere and changes by
means of rotation operator [5]:
\[
R\left( {\theta ,\varphi }\right) =\left( \begin{array}{cc}
\cos \theta  & -ie^{-i\varphi }\sin \theta  \\
-ie^{i\varphi }\sin \theta  & \cos \theta \end{array}\right) .
\]
\noindent For simplicity we shall work with equatorial qubits,
then  $\varphi = \pi /2$ and
\[
\begin{array}{l}
R\left( {\theta }\right) \left| {0}\right\rangle =\cos \theta \left|
{0}\right\rangle +\sin \theta \left| {1}\right\rangle , \\
\\
R\left( {\theta }\right) \left| {1}\right\rangle =-\sin \theta
\left| {0}\right\rangle +\cos \theta \left| {1}\right\rangle .
\end{array}\]
\noindent Pauli matrices can be written down as projectors
\[
\begin{array}{l}
\hat {\mathrm{I}} = \sigma _{0} = \left| {0} \right\rangle
\left\langle {0} \right| + \left| {1} \right\rangle \left\langle {1}
\right|, \quad \sigma _{1} = \left| {1} \right\rangle \left\langle
{0} \right| + \left| {0}
\right\rangle \left\langle {1} \right|,\\
\\
\quad \sigma _{2} = i\left( {\left| {1} \right\rangle \left\langle
{0} \right| - \left| {0} \right\rangle \left\langle {1} \right|}
\right),
 \quad \sigma _{3} = \left| {0}
\right\rangle \left\langle {0} \right| - \left| {1} \right\rangle \left\langle {1}
\right|.
\end{array}
\]
\noindent Action of Pauli matrixes on an input $\left| {\psi
_{0}}\right\rangle $ is given by expression

\begin{equation}
\left| {\psi _{i}}\right\rangle =\sigma _{i}\left| {\psi _{0}}\right\rangle .
\label{1.2}
\end{equation}

\noindent We shall note at once, that scalar product $\left\langle
{{\psi _{0}}}\right| \left. {{\psi _{2}}}\right\rangle =0$. It
means, that for any  $\left| {\psi _{0}}\right\rangle $   with the
help of the universal operator $NOT=-i\sigma _{2}$ it is possible
to create an orthogonal state $\left| {\psi }_{2}\right\rangle $.
For corresponding (\ref{1.2}) matrixes of density we shell receive

\begin{equation}
\rho _{i}^{in}=\left| {\psi _{i}}\right\rangle \left\langle {\psi _{i}}\right| .
\label{1.3}
\end{equation}

\noindent Similarity of two quantum states  $\left| {\psi}
\right\rangle $ and $\left| {\chi} \right\rangle $ is determined
by overlapping of their wave functions [6]:

\begin{equation}
F=\left\langle {\psi }\right| \rho _{\chi }\left| {\psi }\right\rangle , \label{1.4}
\end{equation}

\noindent where $\rho _{\chi }=\left| {\chi }\right\rangle
\left\langle {\chi }\right| $.

The basic two-qubit gate is operation of controlled-NOT (CNOT) [5]:

\begin{equation}
P_{12}\left| {x,y}\right\rangle =\left| {x,x\oplus y}\right\rangle \label{1.5}
\end{equation}

\noindent where $\oplus $ is the modulus-2 summation. The first
qubit in  this expression refers to as control. It does not vary
at CNOT transformation. The second qubit is a  controllable or the
target-qubit of CNOT operator. If indexes of the operator to
change places: $P_{12}=P_{21}$, the second qubit becomes control,
and the first becomes a target qubit.  CNOT operator allows to
create the entangled state of two qubits. Studying and use of the
entangled states is one of the basic problems of quantum
calculations.

\section{Two qubit QCM}

The ideal cloning of the qubit is forbidden. Really, as Wooters
and Zurek have shown [1], if there is such a linear unitary
operator, that
\[
U\left| {\alpha }\right\rangle \left| {\nu }\right\rangle =\left| {\alpha }\right\rangle
\left| {\alpha }\right\rangle ,
\]
\noindent than for orthogonal states  $\left\langle {\alpha
}\right| \left. {\beta }\right\rangle =0$ we have
\[
U\left| {\alpha +\beta }\right\rangle \left| {\nu }\right\rangle =\left| {\alpha +\beta
}\right\rangle \left| {\alpha +\beta }\right\rangle =\left| {\alpha }\right\rangle
\left| {\alpha }\right\rangle +\left| {\beta }\right\rangle \left| {\alpha
}\right\rangle +\left| {\alpha }\right\rangle \left| {\beta }\right\rangle +\left|
{\beta }\right\rangle \left| {\beta }\right\rangle .
\]
\noindent On the other hand, owing to linearity $U$:
\[
U\left| {\alpha +\beta }\right\rangle \left| {\nu }\right\rangle =U\left| {\alpha
}\right\rangle \left| {\nu }\right\rangle +U\left| {\beta }\right\rangle \left| {\nu
}\right\rangle =\left| {\alpha }\right\rangle \left| {\alpha }\right\rangle +\left|
{\beta }\right\rangle \left| {\beta }\right\rangle .
\]
\noindent We came to various expressions that  proves
impossibility of existence of such operation. In the present work
the les complicated  task is considered: whether it is possible to
receive, even nonideal  copy of  any quantum state, and what
quantum operations are necessary for this purpose?

\subsection{One-operational QCM}

The elementary QCM can be constructed from a single CNOT operator
(Fig.1). Submitting on one input of the QCM an arbitrary quantum
state $\left| {\psi _{0}}\right\rangle $, and on other - a qubit
in state $\left| {0}\right\rangle $ we shall receive

\[
\left| {\Psi ^{in}} \right\rangle = \left| {\psi _{0}} \right\rangle \left| {0}
\right\rangle = \alpha \left| {00} \right\rangle + \beta \left| {10} \right\rangle .
\]

\begin{figure}[h]
\centering
  \includegraphics[]{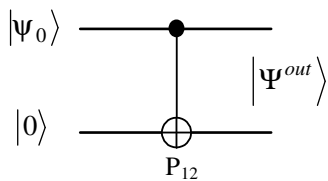}
  \caption{One-operational QCM}
\end{figure}

\noindent The work of the QCM is reduced to action by CNOT operator on
input qubits
\[
\left| {\Psi ^{out}} \right\rangle = P_{12} \left| {\Psi ^{in}}
\right\rangle = \alpha \left| {00} \right\rangle + \beta \left|
{11} \right\rangle ,
\]
\noindent and on an output we receive the entangled state (ebit).
The density matrix of the output ebit is given by expression $\rho
^{out} = \left| {\Psi ^{out}} \right\rangle \left\langle {\Psi
^{out}} \right|$. The reduced density operators of the output
qubits look like

\begin{equation}
\rho _{1,2}^{out}=\alpha ^{2}\left| {0}\right\rangle \left\langle {0}\right| +\beta
^{2}\left| {1}\right\rangle \left\langle {1}\right| .  \label{2.1}
\end{equation}

At the output of the QCM it is received two identical states. It is
necessary to determine, how output states differ from the input
qubit $\left| {\psi _{0}}\right\rangle $. Comparing (\ref{2.1})
and (\ref{1.3}) it is easy to see, that the output density
operator is expressed through the input density operator as
follows:
\[
\rho _{1,2}^{out} = \frac{{1}}{{2}}\rho _{0}^{in} + \frac{{1}}{{2}}\rho _{3}^{in} .
\]
Looking at this expression it is possible to make the assumption,
that the target state on $ 50 \% $ coincides with input state and
has $ 50 \%$ of an impurity. However more detailed consideration
results in other conclusion. Really, scalar product
\[
\left\langle {{\psi _{0}}}\right| \left. {{\psi }_{3}}\right\rangle =\alpha ^{2}-\beta
^{2}\neq 0,
\]
\noindent i.e. states are not orthogonal, and it means, that wave
functions $\left| {\psi _{0}}\right\rangle $ and $\left| {\psi
_{3}}\right\rangle $  are overlapped and  part of the information
concerning $\left| {\psi _{0}}\right\rangle $ is contained in a
density matrix  $\rho _{3}^{in}$. Therefore for calculation of
coppying accuracy we shall take advantage of the formula
(\ref{1.4})

\begin{equation}
F=\left\langle {\psi _{0}}\right| \rho _{1,2}^{out}\left| {\psi _{0}}\right\rangle
=\alpha ^{4}+\beta ^{4}.  \label{2.2}
\end{equation}

From last expression it is visible, that accuracy of cloning
depends on an input state of the original, so received QCM is not
universal. It cannot cloning any, beforehand unknown states with
identical accuracy. Averages (\ref{2.2}) on Bloch sphere we shall
receive:
\[
\bar {F} = \frac{{1}}{{2\pi} }\int\limits_{0}^{2\pi} {Fd\theta} = \frac{{3}}{{4}}.
\]

\subsection{Two-operator QCM}

Now we shall rotate input $\left| {0}\right\rangle $ qubit for
increase of cloning fidelity (Fig.2). Rotation of the $\left|
{0}\right\rangle $ qubit on angle $\phi $ gives
\[
\left| {\psi _{R}}\right\rangle =R\left( {\phi }\right) \left| {0}\right\rangle =\cos
\phi \left| {0}\right\rangle +\sin \phi \left| {1}\right\rangle .
\]
\begin{figure}[h]
\centering
  \includegraphics[]{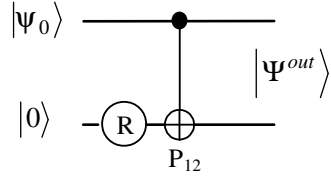}
  \caption{Two-operator QCM}
\end{figure}

\noindent After that, the state  $\left| {\Psi ^{in}}
\right\rangle = \left| {\psi _{0} \psi _{R}} \right\rangle $ comes
to the input channel of the CNOT. At the output of the QCM we'll
receive $\left| {\Psi ^{out}} \right\rangle = P_{12} \left| {\Psi
^{in}} \right\rangle $ with the reduced density matrixes
\[
\rho _{1}^{out}=\alpha ^{2}\left| {0}\right\rangle \left\langle {0}\right| +2\alpha
\beta \cos \phi \sin \phi \left( {\left| {0}\right\rangle \left\langle {1}\right|
+\left| {1}\right\rangle \left\langle {0}\right| }\right) +\beta ^{2}\left|
{1}\right\rangle \left\langle {1}\right| ,
\]

\begin{eqnarray*}
\rho _{2}^{out} &=&\left( {\alpha ^{2}\cos ^{2}\phi +\beta ^{2}\sin
^{2}\phi
}\right) \left| {0}\right\rangle \left\langle {0}\right|  \\
&+&\cos \phi \sin \phi \left( {\left| {0}\right\rangle \left\langle
{1} \right| +\left| {1}\right\rangle \left\langle {0}\right|
}\right) +\left( { \alpha ^{2}\sin ^{2}\phi +\beta ^{2}\cos ^{2}\phi
}\right) \left| {1} \right\rangle \left\langle {1}\right|
\end{eqnarray*}

\noindent and average fidelity
\[
\bar{F}_{1}=\frac{{2}}{{3}}\left( {\cos \phi \sin \phi +1}\right) ,\quad
\bar{F}_{2}=\frac{{\pi }}{{4}}\cos \phi \sin \phi +\frac{{2}}{{3}}\cos ^{2}\phi
+\frac{{1}}{{3}}\sin ^{2}\phi .
\]
\begin{figure}[h]
\centering
  \includegraphics[]{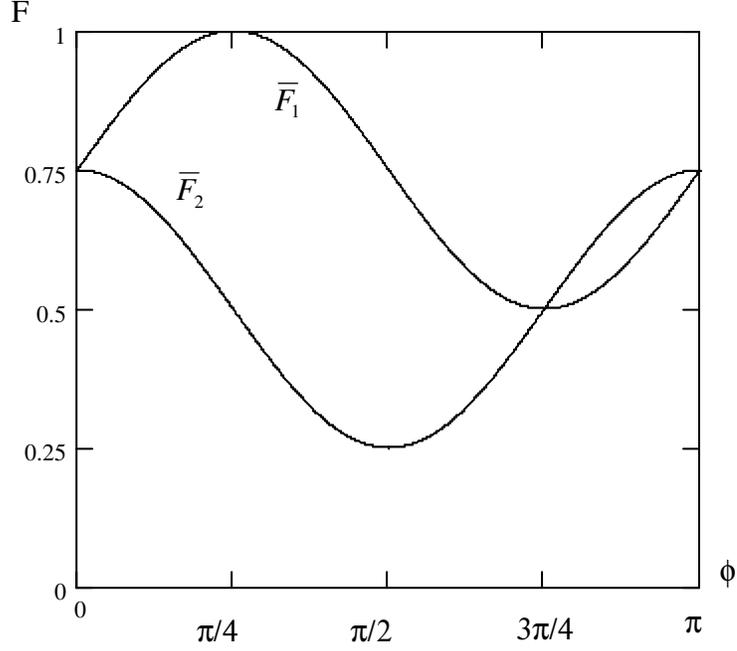}
  \caption{Dependence of the average fidelity $\bar{F}_{1,2}$ on rotation angle $\phi $ for two-operator QCM }
\end{figure}

As well as in previous case, cloning fidelity depends on the
initial state of the qubit-original. In addition, accuracy depends
on a angle  $\phi $ of the rotation $\left| {0}\right\rangle $
qubit. Here it is possible to specify some cases (Fig.3):

1. at  $\phi =0^{\circ }$ we shall receive two copies with
identical average fidelity: $\bar{F}_{1,2}=3/4$.

2. at  $\phi = \pi /4$ the accuracy dispersion of the first qubit
is equal to zero: $\overline {F_{1}^{2}} - \bar {F}_{1}^{2} = 0$
and  the states are separable, i.e.
\[
\left| {\Psi ^{out}}\right\rangle =P_{12}\left| {\psi _{0}}\right\rangle \left| {\psi
_{R}}\right\rangle =\left| {\psi _{0}}\right\rangle \left| {\psi _{R}}\right\rangle
=\left| {\Psi ^{in}}\right\rangle .
\]
3. at  $\phi = \pi /2$ measurement results of output states are
absolutely anticorrelitive:
\[
\frac{{\overline {F_{1} F_{2}} - \bar {F}_{1} \bar {F}_{2}} }{{\sqrt {\overline
{F_{1}^{2}} - \bar {F}_{1}^{2}} \sqrt {\overline {F_{2}^{2}} - \bar {F}_{2}^{2}} } } = -
1
\]
4. at  $\phi =3\pi /2$ average fidelity of output states again
coincides, but we have received "no" results:
$\bar{F}_{1}=\bar{F}_{2}=1/2$.

\section{Tree-qubit QCM}

More complex case of quantum cloning is carried out by Buzek and
Hillery's universal QCM (UQCM BH) [2]. Universality of this
machine is that its cloning accuracy does not depend on a state of
the qubit-original. We'll search for the reduced density matrixes
of output as decomposition on projectors of orthogonal wave
functions   $\left| {\psi _{0}}\right\rangle $ and $\left| {\psi
_{2}}\right\rangle $:

\begin{equation}
\rho _{1,2}^{out}=f_{0}^{2}\rho _{0}^{in}+f_{2}^{2}\rho _{2}^{in}, \label{3.1}
\end{equation}

\noindent where $f_{0}^{2}$ and $f_{2}^{2}$ are so far unknown
factors. Substituting in (\ref{3.1}) decomposition $\rho
_{0}^{in}$ and $\rho _{2}^{in}$ from (\ref{1.3})  we shall receive

\begin{equation}
\rho _{0,1}^{out}=\left( {\alpha ^{2}f_{0}^{2}+\beta
^{2}f_{2}^{2}}\right) \left| {0}\right\rangle \left\langle
{0}\right| +\alpha \beta \left( {f_{0}^{2}-f_{2}^{2}}\right)
\left( {\left| {0}\right\rangle \left\langle {1}\right| +\left|
{1}\right\rangle \left\langle {0}\right| }\right) +\left( {\alpha
^{2}f_{2}^{2}+\beta ^{2}f_{0}^{2}}\right) \left| {1}\right\rangle
\left\langle {1}\right| . \label{3.2}
\end{equation}

\noindent It is easy to show, that reduced density (\ref{3.2}) can
be received from three qubits mix

\begin{equation}
\left| {\Psi ^{out}}\right\rangle =\left| {\Phi _{0}}\right\rangle _{01}\left|
{0}\right\rangle _{2}+\left| {\Phi _{1}}\right\rangle _{01}\left| {1}\right\rangle _{2},
\label{3.3}
\end{equation}

\noindent where
\[
\left| {\Phi _{0}} \right\rangle = \alpha \sqrt {f_{0}^{2} - f_{2}^{2}} \left| {00}
\right\rangle + \beta f_{2} \left| {01} \right\rangle + \beta f_{2} \left| {10}
\right\rangle ,
\]
\[
\left| {\Phi _{1}}\right\rangle =\beta \sqrt{f_{0}^{2}-f_{2}^{2}}\left|
{00}\right\rangle +\alpha f_{2}\left| {01}\right\rangle +\alpha f_{2}\left|
{10}\right\rangle .
\]
\noindent Then,

\begin{eqnarray}
\rho _{0,1}^{out} &=&\left( {\alpha ^{2}f_{0}^{2}+\beta
^{2}f_{2}^{2}}
\right) \left| {0}\right\rangle \left\langle {0}\right| \label{3.4} \\
&+&2\alpha \beta f_{2}\sqrt{f_{0}^{2}-f_{2}^{2}}\left( {\left| {0}
\right\rangle \left\langle {1}\right| +\left| {1}\right\rangle
\left\langle {0}\right| }\right) +\left( {\alpha ^{2}f_{2}^{2}+\beta
^{2}f_{0}^{2}}\right) \left| {1}\right\rangle \left\langle
{1}\right|, \nonumber
\end{eqnarray}

\begin{equation}
\rho _{2}^{out}=\frac{{1}}{{2}}\left( {f_{0}^{2}-f_{2}^{2}}\right) \rho
_{0}^{in}+2f_{2}^{2}\rho _{1}^{in}+\frac{{1}}{{2}}\left( {f_{0}^{2}-f_{2}^{2}}\right)
\rho _{2}^{in}.  \label{3.5}
\end{equation}

\noindent In order cross components in (\ref{3.2}) and (\ref{3.4})
coincided, it is necessary to performe the condition
\[
2f_{2}\sqrt{f_{0}^{2}-f_{2}^{2}}=f_{0}^{2}-f_{2}^{2}
\]
\noindent or

\begin{equation}
f_{0}^{2}=\frac{{5}}{{6}},f_{2}^{2}=\frac{{1}}{{6}}.  \label{3.6}
\end{equation}

\noindent From orthogonality $\left| {\psi _{0}} \right\rangle $
and $\left| {\psi _{2} } \right\rangle $ follows, that the factor
of decomposition $f_{0}^{2} $ and is a cloning fidelity which
coincides with average fidelity.

Sometimes the criterion of universality of quantum cloning is
determined by an opportunity of representation of the output state
as
\[
\rho ^{out} = s\rho ^{in} + \frac{{1 - s}}{{2}}\hat {\mathrm{I}},
\]
\noindent where \textit{s} is scaling factor. It is easy to show,
that the latter expression is a direct consequence from orthogonal
decomposition (\ref{3.1}):
\[
\rho ^{out}=\left( {f_{0}^{2}-f_{2}^{2}}\right) \rho ^{in}+f_{2}^{2}\hat{\mathrm{I}}.
\]
\noindent Hence there follows connection between the dimensional
factor  \textit{s} and factors of decomposition $f_{0}^{2} $ and
$f_{2}^{2} $.

Small increase of fidelity can be obtained, using phase-covariant
(PC) UQCM [3] in which the output state is reduced from three
qubit mix such as:

\begin{equation}
\left| {\Psi ^{out}}\right\rangle =\left| {0}\right\rangle _{0}\left| {\Phi
_{0}}\right\rangle _{12}+\left| {1}\right\rangle _{0}\left| {\Phi _{1}}\right\rangle
_{12},  \label{3.7}
\end{equation}

\noindent where
\[
\left| {\Phi _{0}} \right\rangle = \alpha x\left| {00} \right\rangle + \beta y\left(
{\left| {01} \right\rangle + \left| {10} \right\rangle} \right) + \alpha z\left| {11}
\right\rangle ,
\]
\[
\left| {\Phi _{1}}\right\rangle =\beta z\left| {00}\right\rangle +\alpha y\left( {\left|
{01}\right\rangle +\left| {10}\right\rangle }\right) +\beta x\left| {11}\right\rangle .
\]
\noindent Then,

\begin{eqnarray}
\rho _{1,2}^{out}&=&\left( {\alpha ^{2}\left( {x^{2}+y^{2}}\right)
+\beta ^{2}\left( {y^{2}+z^{2}}\right) }\right) \left|
{0}\right\rangle \left\langle {0}\right| + \label{3.8} \\
&+& 2\alpha \beta \left( {xy + yz} \right)\left( {\left| {0}
\right\rangle \left\langle {1} \right| + \left| {1} \right\rangle
\left\langle {0} \right|} \right) + \left( {\alpha ^{2}\left( {y^{2}
+ z^{2}} \right) + \beta ^{2}\left( {x^{2} + y^{2}} \right)}
\right)\left| {1} \right\rangle \left\langle {1} \right|. \nonumber
\end{eqnarray}

\noindent Comparing (\ref{3.8}) and (\ref{3.2}) we see, that in order to
obtain the reduced output mix such as (\ref{3.1}) it is necessary
to perform the following conditions:

\begin{equation}
x^{2}+y^{2}=f_{0}^{2},  \label{3.9}
\end{equation}

\begin{equation}
y^{2}+z^{2}=f_{2}^{2},  \label{3.10}
\end{equation}

\begin{equation}
2\left( {xy+yz}\right) =f_{0}^{2}-f_{2}^{2},  \label{3.11}
\end{equation}

\begin{equation}
f_{0}^{2}+f_{2}^{2}=1.  \label{3.12}
\end{equation}

\noindent Obtained optimal task with criterion function
(\ref{3.9}) $f_{0}^{2}\rightarrow max$ and restrictions
(\ref{3.10})- (\ref{3.12}) has the solution

\begin{center}
$f_{0}^{2}\left( {max}\right)
=\dfrac{{1}}{{2}}+\dfrac{{1}}{{\sqrt{8}}}, \quad$ at $\quad
x=\dfrac{{1}}{{2}}+\dfrac{{1}}{\sqrt{8}},\quad$
$y=\dfrac{{1}}{\sqrt{8}},\quad$
$z=\dfrac{{1}}{{2}}-\dfrac{{1}}{\sqrt{8}}$.
\end{center}

\noindent Let's notice, that if in system (\ref{3.9})-(\ref{3.12})
we take $z=0$, we shall obtain UQCM BH (\ref{3.3}).

It is obvious, that for the work of the UQCM it is necessary to
have in addition to the qubit-original two more $\left|
{00}\right\rangle $-qubits. It is necessary to find such
transformations of input system of qubits  $\left| {\psi
_{0}}\right\rangle \otimes \left| {0}\right\rangle \otimes \left|
{0}\right\rangle $, which will result to $\left| {\Psi
^{out}}\right\rangle $  (\ref{3.3}) or (\ref{3.7}).

It is convenient to divide the process of cloning into two stages
(Fig.4). At the first stage the entangled state is made from two
$\left| {00} \right\rangle $-qubits with the help of the rotate
and CNOT operators:

\begin{equation}
\left| {00}\right\rangle _{12}\rightarrow \left| {\Psi ^{prep}}\right\rangle
=C_{1}\left| {00}\right\rangle +C_{2}\left| {01}\right\rangle +C_{3}\left|
{10}\right\rangle +C_{4}\left| {11}\right\rangle .  \label{3.13}
\end{equation}

\noindent For UQCM BH it corresponds to the first part
(\ref{3.3}):

\begin{equation}
\left| {00}\right\rangle _{12}\rightarrow \left| {\Psi ^{prep}}\right\rangle
=\sqrt{f_{0}^{2}-f_{2}^{2}}\left| {00}\right\rangle +f_{2}\left| {01}\right\rangle
+f_{2}\left| {10}\right\rangle .  \label{3.14}
\end{equation}

\noindent For UQCM PC, in accordance with  (\ref{3.7}) we'll
receive

\begin{equation}
\left| {00}\right\rangle _{12}\rightarrow \left| {\Psi ^{prep}}\right\rangle
=f_{0}\left| {00}\right\rangle +\sqrt{\frac{{f_{0}^{2}-f_{2}^{2}}}{{2}}}\left( {\left|
{10}\right\rangle +\left| {01}\right\rangle }\right) +f_{2}\left| {11}\right\rangle .
\label{3.15}
\end{equation}

\noindent It is a stage of preparation UQCM for work. At the second stage
entangling of input $\left| {\psi _{0}}\right\rangle $ and
prepared $\left| {\Psi ^{prep}}\right\rangle $ qubits by the CNOT
operators is made: $\left| {\psi _{0}}\right\rangle \otimes \left|
{\Psi ^{prep}}\right\rangle \rightarrow \left| {\Psi
^{out}}\right\rangle $. We shall examine more attentively each of
the stages.

\subsection{Preparation of the UQCM for work}

Process of preparation of the UQCM for work consists in obtaining
of two   $\left| {00}\right\rangle $-qubits of the entangled state
(\ref{3.13}) with the help of rotation operators  $R\left( {\theta
}\right) $ and CNOT:

\begin{eqnarray}
\left| {\Psi ^{prep}}\right\rangle &=&R_{1}\left( {\theta
_{3}}\right) P_{21}R_{2}\left( {\theta _{2}}\right)
P_{12}R_{1}\left( {\theta
_{1}}\right) \left| {00}\right\rangle _{12}  \label{3.16} \\
&=&\left( {\cos \theta _{1}\cos \theta _{2}\cos \theta _{3}+\sin
\theta _{1}\sin \theta
_{2}\sin \theta _{3}}\right) \left| {00}\right\rangle \nonumber \\
&+&\left( {\sin \theta _{1}\cos \theta _{2}\cos \theta _{3}-\cos
\theta
_{1}\sin \theta _{2}\sin \theta _{3}}\right) \left| {01}\right\rangle \nonumber \\
&+&\left( {\cos \theta _{1}\cos \theta _{2}\sin \theta _{3}-\sin
\theta _{1}\sin \theta
_{2}\cos \theta _{3}}\right) \left| {10}\right\rangle \nonumber \\
&+&\left( {\cos \theta _{1}\sin \theta _{2}\cos \theta _{3}+\sin
\theta _{1}\cos \theta _{2}\sin \theta _{3}}\right) \left|
{11}\right \rangle . \nonumber
\end{eqnarray}

\noindent Comparing (\ref{3.13}) and (\ref{3.16}) we'll receive
the system

\begin{equation}
\begin{array}{l}
\cos \theta _{1}\cos \theta _{2}\cos \theta _{3}+\sin \theta _{1}\sin \theta
_{2}\sin \theta _{3}=C_{1}, \\
\sin \theta _{1}\cos \theta _{2}\cos \theta _{3}-\cos \theta _{1}\sin \theta
_{2}\sin \theta _{3}=C_{2}, \\
\cos \theta _{1}\cos \theta _{2}\sin \theta _{3}-\sin \theta _{1}\sin \theta
_{2}\cos \theta _{3}=C_{3}, \\
\cos \theta _{1}\sin \theta _{2}cos\theta _{3}+\sin \theta
_{1}\cos \theta _{2}\sin \theta _{3}=C_{4},\end{array}
\label{3.17}
\end{equation}

\noindent which has the solution

\[
cos^{2}\theta _{1} = \frac{{C_{2}^{2} - C_{3}^{2}} }{{1 - 2C_{3}^{2} - 2C_{4}^{2}} } +
cos^{2}\theta _{3} \frac{{1 - 2C_{2}^{2} - 2C_{4}^{2}} }{{1 - 2C_{3}^{2} - 2C_{4}^{2}}
},
\]
\[
cos^{2}\theta _{2} = \frac{{C_{3}^{2} + C_{4}^{2} - cos^{2}\theta _{3}} }{{1 -
2cos^{2}\theta _{3}} },
\]
\[
cos^{2}\theta _{3}=\frac{{1}}{{2}}\left( {1\pm
\frac{{1-2C_{3}^{2}-2C_{4}^{2}}}{{1-4\left(
{C_{1}^{2}C_{4}^{2}+C_{2}^{2}C_{3}^{2}}\right) }}\sqrt{1-4\left(
{C_{1}^{2}C_{4}^{2}+C_{2}^{2}C_{3}^{2}}\right) +8C_{1}C_{2}C_{3}C_{4}}}\right) .
\]

\noindent Now, substituting as the right part (\ref{3.17}) values
(\ref{3.14}), (\ref{3.15}) we shall have:

\noindent for the UQCM BH

\begin{eqnarray*}
\cos ^{2}\theta _{1} &=&\cos ^{2}\theta _{3}=\frac{1}{2}\left(
1+\frac{1}{\sqrt{2}}\right) , \\
\cos ^{2}\theta _{2} &=&\frac{1}{2}+\frac{\sqrt{2}}{3},
\end{eqnarray*}

\noindent for the UQCM PC

\begin{eqnarray}
\cos ^{2}\theta _{1} &=&\cos ^{2}\theta _{3}=\frac{1}{2}\left(
1+\frac{1}{\sqrt{2}}\right) ,  \label{3.18} \\
\cos ^{2}\theta _{2} &=&1.  \nonumber
\end{eqnarray}

Thus, at the first stage we have received angles of rotation
operators
\[
R_{1}\left( {\theta _{1}}\right) ,R_{2}\left( {\theta _{2}}\right)
,R_{1}\left( {\theta _{3}}\right)
\]

\noindent which together with action of CNOT operators allow to
create state $\left| {\Psi ^{prep}}\right\rangle $ (\ref{3.14}),
(\ref{3.15}) from two $\left| {00}\right\rangle $-qubits. We have
prepared UQCM to work. Actually, the cloning process is carried out
at the second stage.

\subsection{A stage of cloning}

Let's choose such combination of CNOT operators which will make an
output state $\left| {\Psi ^{out}}\right\rangle $ from input
$\left| {\psi _{0}}\right\rangle $ and prepared $\left| {\Psi
^{prep}}\right\rangle $ qubits. We'll consider more in detail the
construction sequence of such operators on the example of UQCM PC.
From state $\left| {\psi _{0}}\right\rangle $ and (\ref{3.15})

\begin{eqnarray}
\left| {\psi _{0}}\right\rangle \otimes \left| {\Psi
^{prep}}\right\rangle &=&\alpha \left( {f_{0}^{2}\left|
{000}\right\rangle +\frac{{f_{0}^{2}-f_{2}^{2}}}{{2}}\left( {\left|
{001}\right\rangle +\left| {010}\right\rangle }\right)
+f_{2}^{2}\left| {011}\right\rangle }\right) \nonumber \\
&+& \beta \left( {f_{2}^{2} \left| {100} \right\rangle +
\frac{{f_{0}^{2} - f_{2}^{2}} }{{2}}\left( {\left| {101}
\right\rangle + \left| {110} \right\rangle} \right) + f_{0}^{2}
\left| {111} \right\rangle} \right) \label{3.19}
\end{eqnarray}

\noindent we are to obtain the output state (\ref{3.7}):

\begin{eqnarray}
\left| {\Psi ^{out}}\right\rangle &=&\alpha \left( {f_{0}^{2}\left|
{000}\right\rangle +\frac{{f_{0}^{2}-f_{2}^{2}}}{{2}}\left( {\left|
{101}\right\rangle +\left| {110}\right\rangle }\right)
+f_{2}^{2}\left| {011}\right\rangle }\right) \nonumber \\
&+&\beta \left( {f_{2}^{2}\left| {100}\right\rangle
+\frac{{f_{0}^{2}-f_{2}^{2}}}{{2}}\left( {\left| {001}\right\rangle
+\left| {010}\right\rangle }\right) +f_{0}^{2}\left|
{111}\right\rangle }\right). \label{3.20}
\end{eqnarray}

\noindent Let's present a state (\ref{3.19}) as

\begin{equation}
\left| {\psi _{0}}\right\rangle \otimes \left| {\Psi ^{prep}}\right\rangle
=C_{xyz}\left| {x,y,z}\right\rangle =C_{xyz}\left| {\Psi ^{in}}\right\rangle _{xyz}
\label{3.21}
\end{equation}

\noindent and output state  (\ref{3.20}) as

\begin{equation}
\left| {\Psi ^{out}}\right\rangle =C_{pqr}\left| {p,q,r}\right\rangle =C_{pqr}\left|
{\Psi ^{out}}\right\rangle _{pqr},  \label{3.22}
\end{equation}

\noindent where $x,y,z,p,q,r = 0,1.$

Comparing values of wave function at identical coefficients
$C_{i}$, we shall reduce transition from (\ref{3.19}) to
(\ref{3.20}) in the table (see Tab.1.)

\begin{center}
Table 1. Truth table for Boolean functions $p\left( {x,y,z}\right)
,q\left( {x,y,z}\right) ,r\left( {x,y,z}\right) $.
\[
  \begin{tabular}{cccccc}
  \hline
  x & y & z & p & q & r \\
  \hline
  0 & 0 & 0 & 0 & 0 & 0 \\
  0 & 0 & 1 & 1 & 0 & 1 \\
  0 & 1 & 0 & 1 & 1 & 0 \\
  0 & 1 & 1 & 0 & 1 & 1 \\
  1 & 0 & 0 & 1 & 0 & 0 \\
  1 & 0 & 1 & 0 & 0 & 1 \\
  1 & 1 & 0 & 0 & 1 & 0 \\
  1 & 1 & 1 & 1 & 1 & 1 \\
  \hline
  \end{tabular}
\]
\end{center}

We have received the table of the validity for Boolean
functions of three variables
$p\left({x,y,z}\right),q\left({x,y,z}\right),r\left({x,y,z}\right)$.
According to the given table we shall write down set of disjunctive normal
forms and present them as Zhegalkin polynoms:

\begin{equation}
\begin{array}{l}
p=\bar{x}\&\bar{y}\&z\vee \bar{x}\&y\&\bar{z}\vee x\&\bar{y}\&\bar{z}\vee
x\&y\&z=x\oplus y\oplus z, \\
q=\bar{x}\&y\&\bar{z}\vee \bar{x}\&y\&z\vee x\&y\&\bar{z}\vee x\&y\&z=y, \\
r=\bar{x}\&\bar{y}\&z\vee \bar{x}\&y\&z\vee x\&\bar{y}\&z\vee x\&y\&z=z.\end{array}
\label{3.23}
\end{equation}

\noindent Using the definition of CNOT operator (\ref{1.5}) from
(\ref{3.23}) we'll obtain
\[
\left| {\Psi ^{out}} \right\rangle _{pqr} = \left| {p,q,r} \right\rangle = \left| {x
\oplus y \oplus z,y,z} \right\rangle = P_{21} P_{31} \left| {x,y,z} \right\rangle =
P_{21} P_{31} \left| {\Psi ^{in}} \right\rangle _{xyz} .
\]
Let's sum up construction UQCM PC (Fig.4). The first stage is
described by the expression (\ref{3.16}). However, according to
(\ref{3.18}) $R_{2}\left( {\theta _{2}}\right) =1$, therefore
(\ref{3.16}) can be copied as
\[
\left| {\Psi ^{prep}} \right\rangle = R_{1} P_{21} P_{12} R_{1} \left| {00}
\right\rangle ,
\]
\noindent where the rotation angle of the operator  $R_{1}$ is
determined in (\ref{3.18}). At the second stage two CNOT operators
entangle the qubit-original $\left| {\psi _{0}}\right\rangle $
with prepared state  $\left| {\Psi ^{prep}}\right\rangle $:
\[
\left| {\Psi ^{out}} \right\rangle = P_{21} P_{31} \left| {\psi _{0}} \right\rangle
\otimes \left| {\Psi ^{prep}} \right\rangle .
\]
\noindent As result we shall receive an output three qubits
density matrix  $\rho _{012}^{out}=\left| {\Psi
^{out}}\right\rangle \left\langle {\Psi ^{out}}\right| $ from
which with accuracy
$f_{0}^{2}=\frac{{1}}{{2}}+\frac{{1}}{\sqrt{8}}\cong 0.854$ two
identical bunches  $\rho _{0}^{in}=\left| {\psi _{0}}\right\rangle
\left\langle {\psi _{0}}\right| $ are reduced.

\begin{figure}
\centering
  \includegraphics[]{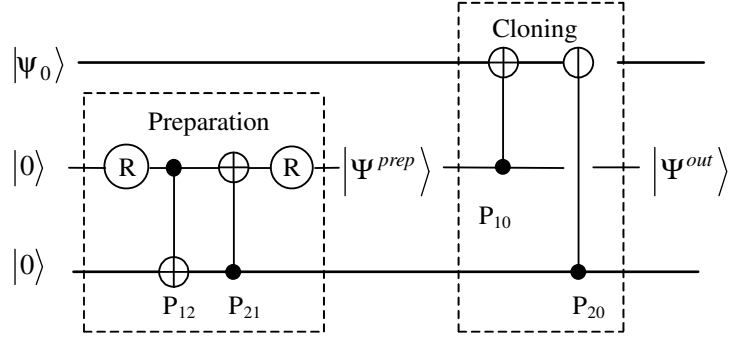}
  \caption{Phase-covariant UQCM. After the preparation procedure it is enough to work on the
qubit-original by two CNOT operators in order to obtain 2 copies with accuracy 0.854.}
\end{figure}

The similar calculation algoritm at comparison of expressions
(\ref{3.3}), (\ref{3.14}) result in sequence of actions of the UQCM BH  [7]:
\[
\left| {\Psi ^{out}} \right\rangle = P_{21} P_{02} P_{10} \left| {\psi _{0} }
\right\rangle \otimes \left| {\Psi ^{prep}} \right\rangle .
\]
We shall note the basic differences of two considered UQCM. For
UQCM BH the pure input state  $\left| {\psi _{0}}\right\rangle $
at the output is registered as a mix (\ref{3.1}). For UQCM PC the
input state  $\left| {\psi _{0}}\right\rangle $ after cloning
procedure has the view
\[
\rho _{0}^{out} = \frac{{3}}{{4}}\rho _{0}^{in} + \frac{{1}}{{4}}\rho _{2}^{in}
\]
\noindent and has already 0.25 parts of the impurity.

The further generalization of calculation algorithms of the UQCM
is connected to the ability of CNOT operator to interchange the
position coefficients $C_{xyz}$ in decomposition (\ref{3.13}).
Then, in order to obtain an output state $\left| {\Psi
^{out}}\right\rangle $ it is necessary to keep only a function
$\left| {\Psi ^{prep}}\right\rangle $ (\ref{3.13}), and it becomes
indifferent where is a factor $C_{xyz}$. As an example we shall
examine UQCM PC (\ref{3.15}). Calculation result are given in
Table 2. The first line of the second column determines the
initial combination of coefficients (\ref{3.15}): $\left(
{C_{1},C_{2},C_{3},C_{4}}\right) =\left(
{\frac{{1}}{{2}}+\frac{{1}}{\sqrt{8}},\frac{{1}}{{\sqrt{8}}},\frac{{1}}{\sqrt{8}},\frac{{1}%
}{{2}}-\frac{{1}}{\sqrt{8}}}\right) $. As $C_{2}=C_{3}$,
rearranging coefficients $C_{i}$ we shall receive 12
various combinations. For each combination on system (\ref{3.17})
we shall find rotation angles $\theta _{1},\theta _{2},\theta
_{3}$, and on algorithm (\ref{3.21})-(\ref{3.23}) - operators of
transition P: $\left| {\psi _{0}}\right\rangle \otimes \left|
{\Psi ^{prep}}\right\rangle \rightarrow \left| {\Psi
^{out}}\right\rangle $. For simplification of record we shall
enter the following approach:
\[
arccos\sqrt{\frac{{1}}{{2}}+\frac{{1}}{\sqrt{8}}}=\frac{{\pi }}{{8}}=22^{0}30^{\prime
},\quad arccos\sqrt{\frac{{1}}{{2}}+\frac{{1}}{\sqrt{6}}}=\frac{{53\pi
}}{{540}}=17^{0}40^{\prime }
\]
\[
arccos\frac{{\sqrt {2 + \sqrt {3}} } }{{2}} = \frac{{\pi} }{{12}} = 15^{0}, \quad
arccos\sqrt {\frac{{1}}{{2}}\left( {1 + \frac{{1}}{\sqrt {3} }} \right)} = \frac{{41\pi}
}{{270}} = 27^{0}20^{\prime}
\]
\noindent For the description of CNOT operator with inversion the
designation is accepted

\[
P_{1\bar {2}} \left| {x,y} \right\rangle = P_{12} \left| {x,\bar {y}} \right\rangle =
\left| {x,x \oplus \bar {y}} \right\rangle = P_{12} \left| {x,\sigma _{1} y}
\right\rangle .
\]
\noindent From table 2 it is visible, that for each $\left| {\Psi
^{prep}}\right\rangle $ there are two ways to obtain $\left| {\Psi
^{out}}\right\rangle $. It is connected to symmetry of an output
state with respect to rearrangement  $\left| {\psi
_{1}^{out}}\right\rangle \leftrightarrow \left| {\psi
_{2}^{out}}\right\rangle $. If we assump that for some
combinations $\left( {C_{1},C_{2},C_{3},C_{4}}\right) $ there are
some decisions $\theta _{1},\theta _{2},\theta _{3}$ it is quite
possible to choose the most simple UQCM from the number of
obtained UQCM (Fig.4).

\newpage

\begin{table}
Table 2. Phase-covariant universal quantum cloning machines.
\bigskip

$\begin{array}{ccccccc} \hline
\\
N & \left( {C_{1},C_{2},C_{3},C_{4}}\right) & \theta _{1} & \theta _{2} & \theta _{3} &
\left| {\Psi ^{out}}\right\rangle _{xyz} & P:\left| {\psi }_{0}\right\rangle \left|
{\Psi ^{prep}}\right\rangle \rightarrow \left|
{\Psi ^{out}}\right\rangle  \\
\\
\hline
\\
 1 & \left( {C_{1},C_{2},C_{2},C_{4}}\right)  & 22^{\circ }30^{\prime } & 0^{\circ
} & 22^{\circ }30^{\prime } &
\begin{array}{c}
\left| x\oplus y\oplus z,y,z\right\rangle  \\
\left| x\oplus y\oplus z,z,y\right\rangle \end{array} &
\begin{array}{c} P_{10}P_{20} \\
P_{12}P_{21}P_{12}P_{10}P_{20}\end{array}
\\
\\
2 & \left( {C_{1},C_{2},C_{4},C_{2}}\right)  & 27^{\circ }20^{\prime } & 15^{\circ } &
17^{\circ }40^{\prime } & \begin{array}{c}
\left| x\oplus z,y,y\oplus z\right\rangle  \\
\left| x\oplus z,y\oplus z,y\right\rangle \end{array} &
\begin{array}{c} P_{12}P_{20} \\ P_{12}P_{21}P_{20}\end{array}
\\
\\
3 & \left( {C_{1},C_{4},C_{2},C_{2}}\right)  & 17^{\circ }40^{\prime } & 15^{\circ } &
27^{\circ }20^{\prime } & \begin{array}{c}
\left| x\oplus y,z,y\oplus z\right\rangle  \\
\left| x\oplus y,y\oplus z,z\right\rangle \end{array} &
\begin{array}{c} P_{21}P_{12}P_{10} \\ P_{21}P_{10}\end{array}
\\
\\
4 & \left( {C_{2},C_{1},C_{2},C_{4}}\right)  & 62^{\circ }40^{\prime } & -15^{\circ } &
17^{\circ }40^{\prime } & \begin{array}{c}
\left| x\oplus \overline{z},y,y\oplus \overline{z}\right\rangle  \\
\left| x\oplus \overline{z},y\oplus \overline{z},y\right\rangle
\end{array} & \begin{array}{c} P_{12}P_{\overline{2}0} \\
P_{12}P_{20}P_{\overline{2}1}\end{array}
\\
\\
5 & \left( {C_{2},C_{1},C_{4},C_{2}}\right)  & 67^{\circ }30^{\prime } & 0^{\circ } &
22^{\circ }30^{\prime } & \begin{array}{c}
\left| x\oplus y\oplus \overline{z},y,\overline{z}\right\rangle  \\
\left| x\oplus y\oplus \overline{z},\overline{z},y\right\rangle
\end{array} & \begin{array}{c}
P_{10}P_{\overline{2}0} \\
P_{21}P_{12}P_{10}P_{21}P_{\overline{2}0}\end{array}
\\
\\
6 & \left( {C_{2},C_{2},C_{1},C_{4}}\right)  & 17^{\circ }40^{\prime } & -15^{\circ } &
62^{\circ }40^{\prime } & \begin{array}{c}
\left| x\oplus \overline{y},z,\overline{y}\oplus z\right\rangle  \\
\left| x\oplus \overline{y},\overline{y}\oplus z,z\right\rangle
\end{array} & \begin{array}{c} P_{21}P_{12}P_{\overline{1}0} \\
P_{10}P_{\overline{2}0}\end{array}
\\
\\
7 & \left( {C_{2},C_{2},C_{4},C_{1}}\right)  & -17^{\circ }40^{\prime } & 75^{\circ } &
-27^{\circ }20^{\prime } & \begin{array}{c}
\left| x\oplus \overline{y},\overline{z},y\oplus z\right\rangle  \\
\left| x\oplus \overline{y},y\oplus z,\overline{z}\right\rangle
\end{array} & \begin{array}{c}
P_{21}P_{12}P_{\overline{1}0} \\
P_{\overline{2}1}P_{\overline{1}0}\end{array}
\\
\\
8 & \left( {C_{2},C_{4},C_{1},C_{2}}\right)  & 22^{\circ }30^{\prime } & 0^{\circ } &
67^{\circ }30^{\prime } & \begin{array}{c}
\left| x\oplus \overline{y}\oplus z,z,\overline{y}\right\rangle  \\
\left| x\oplus \overline{y}\oplus z,\overline{y},z\right\rangle
\end{array} & \begin{array}{c}
P_{12}P_{21}P_{12}P_{\overline{1}0}P_{20} \\
P_{\overline{1}0}P_{20}\end{array}
\\
\\
9 & \left( {C_{2},C_{4},C_{2},C_{1}}\right)  & -27^{\circ }20^{\prime } & 75^{\circ } &
-17^{\circ }40^{\prime } & \begin{array}{c}
\left| x\oplus \overline{z},y\oplus z,\overline{y}\right\rangle  \\
\left| x\oplus \overline{z},\overline{y},y\oplus z\right\rangle
\end{array} & \begin{array}{c}
P_{12}P_{\overline{2}0}P_{21} \\
P_{\overline{1}2}P_{\overline{2}0}\end{array}
\\
\\
10 & \left( {C_{4},C_{2},C_{2},C_{1}}\right)  & 67^{\circ }30^{\prime } & 0^{\circ } &
67^{\circ }30^{\prime } & \begin{array}{c}
\left| x\oplus y\oplus z,\overline{z},\overline{y}\right\rangle  \\
\left| x\oplus y\oplus z,\overline{y},\overline{z}\right\rangle
\end{array} & \begin{array}{c}
P_{12}P_{21}P_{12}P_{\overline{1}0}P_{\overline{2}0} \\
P_{\overline{1}0}P_{\overline{2}0}\end{array}
\\
\\
11 & \left( {C_{4},C_{2},C_{1},C_{2}}\right)  & 27^{\circ }20^{\prime } & -15^{\circ } &
72^{\circ }20^{\prime } & \begin{array}{c}
\left| x\oplus z,\overline{y}\oplus z,\overline{y}\right\rangle  \\
\left| x\oplus z,\overline{y},\overline{y}\oplus z\right\rangle
\end{array} & \begin{array}{c} P_{\overline{1}2}P_{21}P_{20} \\
P_{\overline{1}2}P_{20}\end{array}
\\
\\
12 & \left( {C_{4},C_{1},C_{2},C_{2}}\right)  & 72^{\circ }20^{\prime } & -15^{\circ } &
27^{\circ }20^{\prime } & \begin{array}{c}
\left| x\oplus y,y\oplus \overline{z},\overline{z}\right\rangle  \\
\left| x\oplus y,\overline{z},y\oplus \overline{z}\right\rangle
\end{array} & \begin{array}{c}
P_{\overline{2}1}P_{10} \\ P_{\overline{2}1}P_{12}P_{10}\end{array}\\
\\
\hline
\end{array}$

\end{table}

\end{document}